\documentclass[aps,prl,showpacs,,showkeys,twocolumn]{revtex4}
\usepackage{mathrsfs}
\usepackage{amssymb}
\usepackage{amsmath}
\usepackage{bm}
\usepackage{graphics}
\usepackage{natbib}

\begin{document}
\title{Influences of electron-phonon interaction on quantum transport through one quantum-dot system with side-coupled Majorana zero mode}

\author{Xiao-Qi Wang$^1$}
\author{B. H. Wu$^2$}
\author{Shu-Feng Zhang$^3$}
\author{Qi Wang$^1$}
\author{Wei-Jiang Gong$^{1}$}\email[Corresponding author.
Fax: +086-024-8367-6883; phone: +086-024-8367-8327; Email address: ]
{gwj@mail.neu.edu.cn}

\affiliation{ 1. College of Sciences, Northeastern University,
Shenyang 110819, China \\
2. Department of Applied Physics, Donghua University, Shanghai 201620, China\\
3. School of Physics and Technology, University of Jinan, Jinan, Shandong 250022, China}

\date{\today}

\begin{abstract}
We investigate the influences of the electron-phonon interaction on the transport properties of one quantum-dot system with a side-coupled Majorana zero mode (MZM). Our calculation results show that at the zero-temperature limit, the MZM-governed zero-bias conductance value can be magnified, dependent on the interplay between electron-phonon interaction and dot-MZM coupling. In the case of finite temperature, the electron-phonon interaction makes leading contributions to the suppression of the magnitude of zero-bias conductance, but the effect is different from the case of electron tunneling without MZM. We believe that this work can be helpful for understanding the signature of the MZM in electron transport through mesoscopic circuits.
\end{abstract}
 \maketitle

\bigskip

\section{Introduction}
Following the successful fabrication of graphene, materials with topological nature have attracted enormous attentions in the field of condensed matter physics\cite{a1}. And topological insulators, semimetals, and superconductors have been achieved in experiments, accompanied by the clarification of their leading physics properties. It is well-known that the edges of topological superconductors host localized zero-energy excitations that are commonly referred
to as Majorana zero modes (MZMs)\cite{a2}. These zero modes can be used to implement
the fault-tolerance topological quantum computation due to their non-Abelian
statistical characteristics\cite{quantumC1,quantumC2,Stern}.
During the past years, these favorable characteristics have motivated intense searches for systems which
can achieve such excitations. The first candidate is the superconducted one-dimensional semiconducting
nanowires due to proximity effect\cite{Kitaev,Fu1,Oreg,Sarma}, which have then been realized experimentally with the appearance of MZM's signature\cite{exper1,exper2,Franz}. After then, lots of other schemes have been proposed to generate the MZMs. One is to realize the MZMs in the ferromagnetic quantum chain by laying it on the superconductor surface\cite{ferromagnetic}. The other scheme is to build the vertex of the a topological superconductor of the Bi$_2$Te$_3$/NbSe$_2$ heterostructure\cite{jia1,jia2}. Recently, the MZM has been observed inside a vortex core on the superconducting Dirac surface state of the iron-based superconductor FeTe$_{0.55}$Se$_{0.45}$\cite{science}.
\par
The experimental reports inspire theoretical researchers dedicated themselves to the further investigation about the physics properties of MZMs. And a large number
of interesting phenomena have been demonstrated, such as the resonant
Andreev reflection\cite{Ng1,flensberg}, the special crossed Andreev reflection\cite{Beenakker1}, the fractional
Josephson effect\cite{Josephson1,Josephson2,Josephson3,Josephson4,Josephson5,New1,New2,Fu2}, as well as the nonlocality of MZMs\cite{Zhang}. Based on these results, new proposals have been suggested from different aspects to differentiate the signals of MZMs from the peaks of the quantum transport spectra. Moreover, MZMs
are considered to couple with the regular bound state formed by one quantum dot (QD) to study the interplay between these two bound states\cite{Flensberg0,flensberg2,Lee1,Shen,PRL,Li,Liude,Vernek,Lopez}. For instance, if one QD is considered to insert in the Andreev-transport circuits based on the coupling between metal lead and topological superconductor, the nonlocality of the MZMs can be well observed. It has also been reported that when the QD-based circuit is laterally coupled with one MZM,
the conductance through the QD is influenced by inducing the sharp
decrease of the conductance by a factor of $1\over2$\cite{Liude,Vernek}.
Similar results can be observed when the QD is in the Kondo regime.
Namely, the QD-MZM coupling reduces the unitary-limit value of the
linear conductance by exactly a factor ${3\over4}$\cite{Lopez}. These results
have also been viewed as promising ideas for detecting the MZMs in experiments.
\par
Despite the existed works to describe the properties of MZMs, the experimental observations exhibit difference from the theoretical expectations, even if the simplest resonant Andreev reflection\cite{Argue}. One important reason is that decoherence effects, Andreev bound states, or the additional factors make their contributions to the formation of MZMs, as well as the MZM-assisted quantum transport processes. In view of these facts, researchers have begun to pay attention to the detailed roles of decoherence effects in modulating the MZM-assisted transport behaviors\cite{Sun,Dassama}. However, these existed investigations are not enough to present the comprehensive influences of decoherence mechanisms. Also, it is known that the decoherence effects, especially the electron-phonon (e-ph) interaction, play important roles in modulating the conventional Andreev reflection and Josephson current\cite{Wu1,Wu2,Wu3}. Therefore, in the present work, we would like to investigate the influence of such a typical decoherence effect, i.e., the e-ph coupling, on the transport properties of one QD system with a side-coupled MZM, by assuming it to occur in the QD region. After calculation, we see that at the zero-temperature limit, the zero-bias conductance value can be magnified by the e-ph interaction, differently from the case of zero MZM. This is exactly dependent on the interplay between e-ph interaction and QD-MZM coupling. Instead, at the case of finite temperature, the e-ph interaction makes contributions to the suppression of the magnitude of zero-bias conductance. We believe that this work can be helpful for understanding the signature of the MZM in the presence of decoherence.

 \begin{figure}
 \begin{center}\scalebox{0.45}{\includegraphics{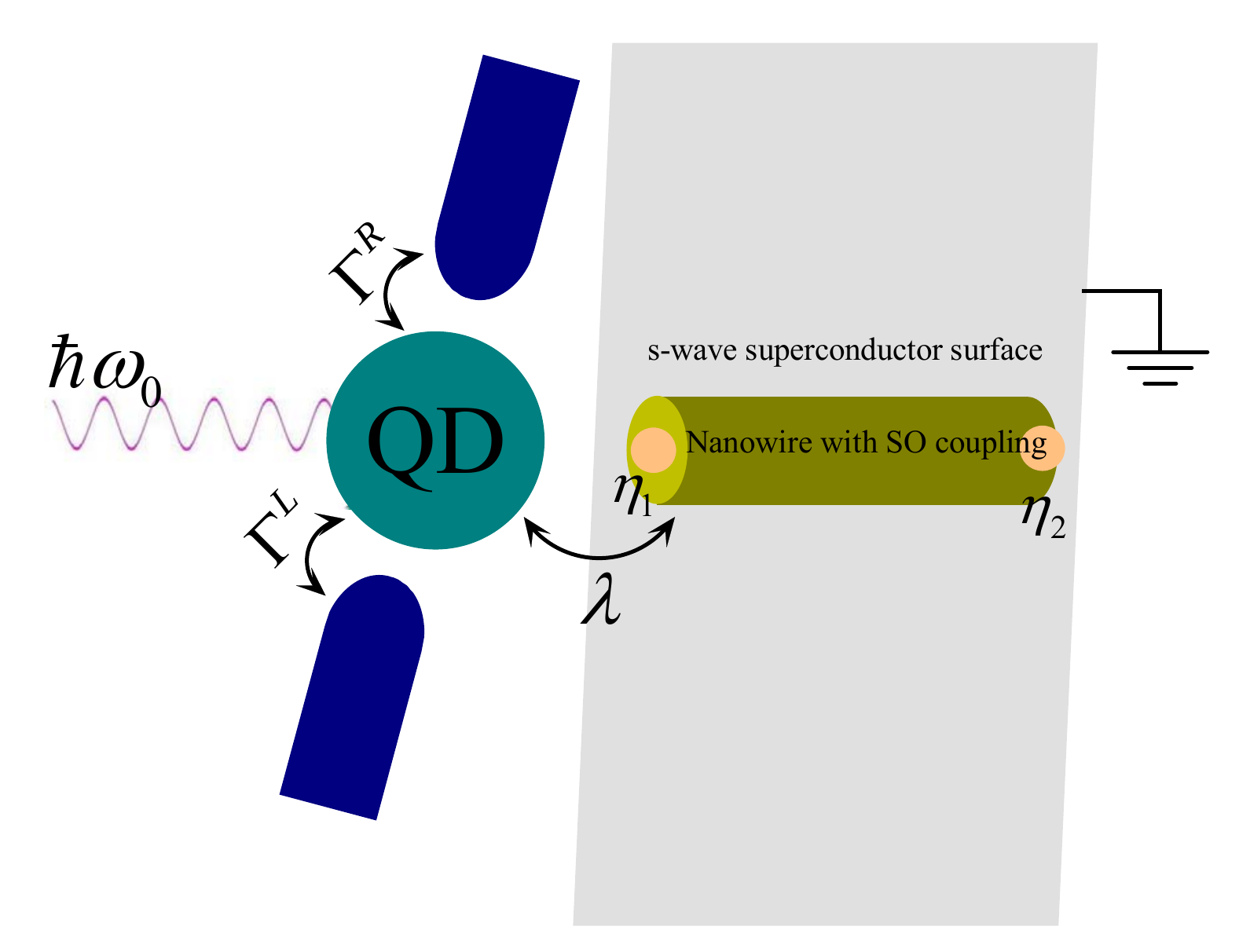}}
 \caption{Illustration of one single-QD system with coupled Majorana bound states. The e-ph interaction takes place in QD region. The Majorana bound states (MBSs) are supposed to form at the ends of the nanowire with spin-orbit coupling which adheres to the $s$-wave superconductor. The MBSs are labeled as $\eta_1$ and $\eta_2$, respectively.} \label{Struct}
  \end{center}
 \end{figure}

\section{Model and formulation\label{theory}}
Our considered QD structure with e-ph
interaction is illustrated in Fig.1. Two leads couple via the QD which suffers from the local e-ph interaction. Besides, one Majorana bound state (MBS) is considered to couple to the QD laterally. Its Hamiltonian can be written as $H=H_{\rm leads}+H_{\rm ph}+H_M+H_D+H_T$.
\par
The first two terms represent the noninteracting electron gas
in the leads and the phonon mode, respectively. They take the forms as
\begin{eqnarray}
H_{\rm leads}&=&\underset{\alpha k}{\sum }\varepsilon
_{\alpha k}c_{\alpha k}^\dag c_{\alpha k}, H_{\rm
ph}=\omega_0a^\dag a.
\end{eqnarray}
$c_{\alpha k}^\dag$ $( c_{\alpha k})$ is the operator to
create (annihilate) an electron in state $|k\rangle$ of lead-$\alpha$ ($\alpha=L,R$). $\varepsilon_{\alpha k}$ is the corresponding level. $\omega_0$ is the frequency of the single-phonon mode (Einstein model), and $a^{\dag}$ $(a)$ is the
phonon creation (annihilation) operator. According to the previous works, such an assumption is feasible, since the LO-mode phonon makes the leading contribution in the QD\cite{AA,AB,AC}. $H_M$ denotes the Hamiltonian of coupled MBSs. The low-energy effective form reads
\begin{eqnarray}
H_{M}=i\epsilon_M\eta_1\eta_2,\label{3}
\end{eqnarray}
which describes the paired MBSs generated at the ends of the nanowire
and coupled to each other by an energy $\epsilon_M\sim e^{-l/\xi}$,
with $l$ being the wire length and $\xi$ the superconducting coherent
length\cite{MBSe}. Next,
\begin{eqnarray}
H_D&=&[\varepsilon_d+\gamma(a^\dag+a)] n_d.
\end{eqnarray}
$n_d=d^\dag d_{}$ and $d^{\dag}$
$(d_{})$ is the electron creation (annihilation) operators of
QD, and $\varepsilon_d$ is the single-energy level of the QD.
$\gamma$ is the coupling constant between the QD electron and
phonon mode.
The last term describes the couplings of the QD to the MBS and the
leads:
\begin{eqnarray}
H_T&=&(\lambda d-\lambda^* d^\dag)\eta_1+\underset{
\alpha k}{\sum }V_{\alpha k} c^\dag_{\alpha
k}d_{}+\mathrm {h.c.}.
\end{eqnarray}
$\lambda$ is the coupling coefficient between the QD and MBS. $V_{\alpha k}$ denotes the coupling between the QD and the
leads, whose magnitude is determined by the overlap of the wavefunctions in the QD and leads.
\par
\par
In the presence of bias voltage applied between the leads, the chemical potentials of them can be written as $\mu_L=\varepsilon_F+{eV_b\over2}$ and
$\mu_R=\varepsilon_F-{eV_b\over2}$ ($\varepsilon_F$ is the Fermi level at equilibrium and can be assumed to be zero). And then, the quantum transport through this system will be driven. The current flow in
lead-$\alpha$ can be evaluated by means of the nonequilibrium Green function
technique\cite{Meir}. Via a straightforward derivation, we obtain the
expression of the current in one lead, e.g., lead-$L$\cite{gongprb}:
\begin{equation}
J_L={e\over h}\int d\omega[
T_{ee}^{LR}(\omega)(f^L_e-f^{R}_e)+
T_{eh}^{LL}(\omega)(f^L_e-f^{L}_h)].\label{dd}
\end{equation}
$f^\alpha_e$ and $f^\alpha_h$ are the Fermi
distributions of the electron and hole in lead-$\alpha$,
respectively. $T_{ee}^{\alpha\alpha'}(\omega)={\rm
Tr}[\Gamma_e^\alpha G^R\Gamma_e^{\alpha'}G^A]$ and
$T_{eh}^{\alpha\alpha}(\omega)={\rm
Tr}[\Gamma_e^\alpha G^R\Gamma_h^{\alpha}G^A]$,
where $G^R$ and $G^A$ are the related and advanced
Green functions. $\Gamma^{\alpha}_{e}$ denotes the coupling matrix between the QD and leads and $\Gamma^{\alpha}_{h}$ is its hole counterpart, which are defined by $\Gamma^{\alpha}_{e(h)}=2\pi\sum_k|V_{\alpha k}|^2\delta(\omega\mp\varepsilon_{\alpha k})$. Within the wide-band limit approximation, there will be
$\Gamma^\alpha_e=\Gamma^\alpha_h=\Gamma^\alpha$. Moreover, in the
symmetric-coupling case where $\Gamma^\alpha=\Gamma$, we can simplify the current formula in this structure
as
\begin{equation}
J={e\over 2h}\int d\omega \Gamma A_d(\omega)(f^L_e-f^R_e),\label{simple}
\end{equation}
with $A_d(\omega)={i}(G^R_d-G^A_d)={i}(G^>_d-G^<_d)$. $G^>_d$ and $G^<_d$ are the greater and lesser Green functions, respectively.
\par
The next step for calculating the current is to figure out the Green functions involved. Due to the existence of e-ph interaction, the Hamiltonian should be managed first.
In addition to the Born approximation, one feasible method is the well-known canonical transformation\cite{can0,can1,can2,can3, Zhu}, which is performed as ${\cal H}=e^SHe^{-S}$ by defining
$S={\gamma\over\omega_0}d^\dag d(a^\dag-a)$. Via this transformation, the Hamiltonian can be divided into two independent parts, i.e, $\mathcal{H}={\cal H}_{\rm
el}+H_{\rm ph}$. And the e-ph interaction can be discussed by only focusing on Hamiltonian of the dressed electron, i.e., ${\cal H}_{\rm
el}$. It is expressed as ${\cal H}_{\rm el}=\underset{\alpha k
}{\sum }\varepsilon _{\alpha k}c_{\alpha k}^\dag
c_{\alpha k}+\tilde{\varepsilon}_d
\tilde{d}^\dag\tilde{d}+(\tilde{\lambda} \tilde{d}-\tilde{\lambda}^* \tilde{d}^\dag)\eta_1+\underset{
\alpha k}{\sum }\tilde{V}_{\alpha k} c^\dag_{\alpha
k}\tilde{d}+\mathrm {h.c.}$ with
$\tilde{\varepsilon}_d=\varepsilon_d-g\omega_0$ and $g={\gamma^2\over\omega^2_0}$. $\tilde{\lambda}=\lambda X$ and
$\tilde{V}_{\alpha k}=V_{\alpha k} X$ are the dressed MZM-QD and QD-lead couplings
with $X=\exp{[-{\gamma\over\omega_0}}(a^\dag-a)]$. One can then understand that the e-ph interaction is manifested as the shifted QD levels and the QD-MZM(lead) couplings. $X$ arises from the canonical transformation
of the particle operator $e^Sde^{-S}=\tilde{d}X$. This reveals that
the interaction between the electron in the QD and the phonon
mode results in an effective phonon-mediated coupling
between the QD and leads. As in dealing with the
localized phonon mode in this work, it is reasonable to replace the operator
$X$ with its expectation value, i.e., $\langle X\rangle=\exp[-g(N_{\mathrm{ph}}+{1\over 2})]$, where $N_{\mathrm{ph}}$ is the phonon population, and can be expressed as $N_{\mathrm{ph}}=[\exp(\beta\omega_0)-1]^{-1}$ with $\beta=1/k_BT$. Meanwhile, as the
operator $X$ is replaced by its expectation value, the interacting
lesser Green function may be separated
\begin{eqnarray}
G^{<}_{d}(t)&=&i\langle d^\dag d(t)
\rangle=i\langle d^\dag e^{i{\cal H}_{\rm el}t}d
e^{-i{\cal H}_{\rm el}t}\rangle\notag\\
 &&\langle X^\dag e^{i H_{\rm ph}t}X
e^{-i H_{\rm ph}t} \rangle \notag\\
&=&\tilde{G}^{<}_{d}(t) e^{-\Phi(-t)},
\end{eqnarray}
and similarly $G^>_d(t)=-i\langle d(t)d\rangle=\tilde{G}^{>}_{d}(t) e^{-\Phi(t)}$.
$\tilde{G}^{<(>)}_{d}(t)$ is the lesser (greater) Green function
for the electron governed by ${\cal H}_{\mathrm{el}}$. The factor $
e^{-\Phi(\mp t)}$ originates from the trace of the phonon parts $\langle X^\dag X(t)\rangle_{\rm ph}$ or $\langle X(t)X^\dag\rangle_{\rm ph}$, in which
$\Phi(t)=g[N_{\mathrm{ph}}(1-e^{i\omega_0
t})+(N_{\mathrm{ph}}+1)(1-e^{-i\omega_0 t})]$.
\par
Using the identity $e^{-\Phi(-t)}=\sum^{\infty}_{l=-\infty}{\cal
{L}}_{l}e^{il\omega_0t}$, the greater and lesser Green functions can
be respectively expanded as
\begin{eqnarray}
&&G^{>}_{d}(\omega)=\sum^{\infty}_{l=-\infty}{\cal
{L}}_{l}\tilde{G}^{>}_{d}(\omega-l\omega_0)\notag\\
&&G^{<}_{d}(\omega)=\sum^{\infty}_{l=-\infty}{\cal
{L}}_{l}\tilde{G}^{<}_{d}(\omega+l\omega_0),
\end{eqnarray}
where the index $l$ stands for the number of phonons involved
and $\mathcal{L}_l$ are the coefficients, depending on temperature and the
strength of e-ph interaction. At finite temperature, ${\cal
{L}}_l=e^{-g(2N_{\mathrm{ph}}+1)}e^{l\omega_0\beta/2}J_{l}(2g\sqrt{N_{\mathrm{ph}}(N_{\mathrm{ph}}+1)})$,
where $J_{l}$ is the complex-argument Bessel function. And at zero temperature, ${\cal {L}}_l$ is
given by
\begin{equation}
{\cal {L}}_l=\left \{
\begin{array}{ccc}
e^{-g}g^l/l!  &&(l\geq0),\\
0  &&(l<0).
\end{array}\right.
\end{equation}
Thus the spectral function can be expressed as
\begin{equation}
A_d(\omega)=\sum_{l=-\infty}^{\infty}i{\cal L}_l[\tilde{G}^>_d(\omega-l\omega_0)-\tilde{G}^<_d(\omega+l\omega_0)].\label{Spectral}
\end{equation}
\par
By means of the equation of motion approach, the retarded Green
function for the dressed electron and hole can be evaluated as
\begin{small}
\begin{eqnarray}
\tilde{G}^R_d(\omega)=\left[\begin{array}{cccc}
 \tilde{g}_{e}(z)^{-1} &0&\tilde{\lambda}^*&0\\
 0&\tilde{g}_{h}(z)^{-1}&-\tilde{\lambda}&0\\
\tilde{\lambda}&-\tilde{\lambda}^*&g_{M}(z)^{-1}&-i\epsilon_M\\
0&0&i\epsilon_M&g_{M}(z)^{-1}
\end{array}\right]^{-1}.\ \notag
\end{eqnarray}
\end{small}
In the above equation,
$\tilde{g}_{e(h)}(z)^{-1}=\omega\mp\tilde{\varepsilon}_d+i\tilde{\Gamma}$ and
$\tilde{g}_{M}(z)^{-1}=\omega+i0^+$ where $\tilde{\Gamma}={1\over2}[\tilde{\Gamma}^L+\tilde{\Gamma}^R]$ with $\tilde{\Gamma}^\alpha=2\pi\sum_k|\tilde{V}_{\alpha k}|^2\delta(\omega-\varepsilon_{\alpha k})$.
Following the Keldysh equation for the Green's functions, i.e., $\tilde{G}^{<(>)}_d=\tilde{G}^R_d\tilde{\Sigma}^{<(>)}\tilde{G}^A_d$ with $\tilde{\Sigma}^<_{e(h)}=i[\tilde{\Gamma}^Lf_{e(h)}^L(\omega)+\tilde{\Gamma}^Rf_{e(h)}^R(\omega)]$ and $\tilde{\Sigma}^>_{e(h)}=i\{\tilde{\Gamma}^L[f_{e(h)}^L(\omega)-1]+\tilde{\Gamma}^R[f_{e(h)}^R(\omega)-1]\}$, one can obtain the dressed greater
and lesser Green's functions:
\begin{eqnarray}
\tilde{G}^{<(>)}_d=|\tilde{G}^R_{d,ee}|^2\tilde{\Sigma}^{<(>)}_{e}+|\tilde{G}^R_{d,eh}|^2\tilde{\Sigma}^{<(>)}_{h}.
\end{eqnarray}
For the case of MZM, i.e., $\epsilon_M=0$, the Green function matrix will be simplified.
And then, $\tilde{G}^R_{d,ee}$ and $\tilde{G}^R_{d,eh}$ can be analytically expressed in the following way:
\begin{small}
\begin{eqnarray}
&&\tilde{G}^R_{d,ee}={\omega(\omega+\tilde{\varepsilon}_d+i\tilde{\Gamma})-|\tilde{\lambda}|^2\over \omega(\omega-\tilde{\varepsilon}_d+i\tilde{\Gamma})(\omega+\tilde{\varepsilon}_d+i\tilde{\Gamma})-2|\tilde{\lambda}|^2(\omega+i\tilde{\Gamma})},\notag\\
&&\tilde{G}^R_{d,eh}={-|\tilde{\lambda}|^2\over \omega(\omega-\tilde{\varepsilon}_d+i\tilde{\Gamma})(\omega+\tilde{\varepsilon}_d+i\tilde{\Gamma})-2|\tilde{\lambda}|^2(\omega+i\tilde{\Gamma})}. 
\end{eqnarray}
\end{small}
Based on these results, the influence of e-ph interaction on the MZM-assisted transport behaviors can be well evaluated, by writing out the current expression:
\begin{eqnarray}
&J&={e\over 2h}\sum_{l=-\infty}^\infty
{\cal L}_l\Gamma \int d\omega [f^L_e(\omega)-f^R_e(\omega)]\times\notag\\
&&\{\tilde{\varrho}(\omega+l\omega_0)[\tilde{\Gamma}^Lf^L_e(\omega+l\omega_0)+\tilde{\Gamma}^Rf^R_e(\omega+l\omega_0)]\notag\\
&-&\tilde{\varrho}(\omega-l\omega_0)[\tilde{\Gamma}^Lf^L_e(\omega-l\omega_0)+\tilde{\Gamma}^Rf^R_e(\omega-l\omega_0)]\notag\\
&+&[|\tilde{G}^R_{d,ee}(\omega-l\omega_0)|^2+|\tilde{G}^R_{d,eh}(\omega+l\omega_0)|^2](\tilde{\Gamma}^L+\tilde{\Gamma}^R)\}
,\notag
\end{eqnarray}
with $\tilde{\varrho}(\omega)=|\tilde{G}^R_{d,ee}(\omega)|^2-|\tilde{G}^R_{d,eh}(\omega)|^2$.
In the case of symmetric QD-lead coupling with $\tilde{\Gamma}^\alpha=\tilde{\Gamma}$, it will be simplified as
\begin{eqnarray}
&J&={e\over 2h}\sum_{l=-\infty}^\infty
{\cal L}_l\Gamma \tilde{\Gamma}\int d\omega [f^L_e(\omega)-f^R_e(\omega)]\times\notag\\
&&\{\tilde{\varrho}(\omega+l\omega_0)[f^L_e(\omega+l\omega_0)+f^R_e(\omega+l\omega_0)]\notag\\
&+&\tilde{\varrho}(\omega-l\omega_0)[2-f^L_e(\omega-l\omega_0)-f^R_e(\omega-l\omega_0)]\notag\\
&+&2[|\tilde{G}^R_{d,eh}(\omega-l\omega_0)|^2+|\tilde{G}^R_{d,eh}(\omega+l\omega_0)|^2]\}.
\end{eqnarray}
\par
In this work, we would like to discuss the influence of e-ph interaction in two cases, i.e., zero and nonzero temperatures, respectively. Accordingly, the zero-temperature differential conductance can be expressed as
\begin{eqnarray}
\mathcal{G}&=&{e^2\over 4h}\sum_{l=0}^\infty
{\cal L}_l\Gamma\tilde{\Gamma} \{\theta(eV_b-l\omega_0)[\tilde{\varrho}(\mu_L)+\tilde{\varrho}(\mu_R)\notag\\
&+&\tilde{\varrho}(\mu_L-l\omega_0)+\tilde{\varrho}(\mu_R+l\omega_0)]+\theta(-eV_b-l\omega_0)\notag\\
&&[\tilde{\varrho}(\mu_L)+\tilde{\varrho}(\mu_R)+\tilde{\varrho}(\mu_L+l\omega_0)+\tilde{\varrho}(\mu_R-l\omega_0)]\notag\\
&+&2[|\tilde{G}^R_{d,eh}(\mu_L-l\omega_0)|^2+|\tilde{G}^R_{d,eh}(\mu_L+l\omega_0)|^2\notag\\
&+&|\tilde{G}^R_{d,eh}(\mu_R-l\omega_0)|^2+|\tilde{G}^R_{d,eh}(\mu_R+l\omega_0)|^2]
\},
\end{eqnarray}
because the Fermi distribution function is reduced to the step function in this case. Alternatively, in the case of finite temperature, the differential conductance will transform into
\begin{eqnarray}
{\cal G}&=&{\beta e^2\over 4h}\sum_{l=-\infty}^\infty
{\cal L}_l\Gamma \tilde{\Gamma}\int d\omega [\mathcal{A}_l(\omega)\tilde{\varrho}(\omega-l\omega_0)\notag\\
&&+2\mathcal{B}(\omega)|\tilde{G}^R_{d,eh}(\omega-l\omega_0)|^2],
\end{eqnarray}
after employing the relationship that ${\cal L}_{-l}=e^{-l\beta\omega_0}{\cal L}_{l}$ and ${\partial f^{L(R)}_e\over \partial V_b}=\pm {e\over2}\beta f^{L(R)}_e[1-f^{L(R)}_e]$. In this formula,
\begin{small}
\begin{eqnarray}
&&\mathcal{A}_l(\omega)=\{f^L_e(\omega)[1-f^L_e(\omega)]+f^R_e(\omega)[1-f^R_e(\omega)]\}\cdot\notag\\
&&\{2+[e^{-l\beta\omega_0}-1][f^L_e(\omega-l\omega_0)+f^R_e(\omega-l\omega_0)]\}\notag\\
&&+[e^{-l\beta\omega_0}-1][f^L_e(\omega)-f^R_e(\omega)]\cdot\{f^L_e(\omega-l\omega_0)\notag\\
&&[1-f^L_e(\omega-l\omega_0)]-f^R_e(\omega-l\omega_0)[1-f^R_e(\omega-l\omega_0)]\},\notag\\
&&\mathcal{B}_l(\omega)=\{f^L_e(\omega)[1-f^L_e(\omega)]+f^R_e(\omega)[1-f^R_e(\omega)]\}\cdot\notag\\
&&[e^{-l\beta\omega_0}+1].\notag
\end{eqnarray}
\end{small}
\begin{figure}
\begin{center}\scalebox{0.24}{\includegraphics{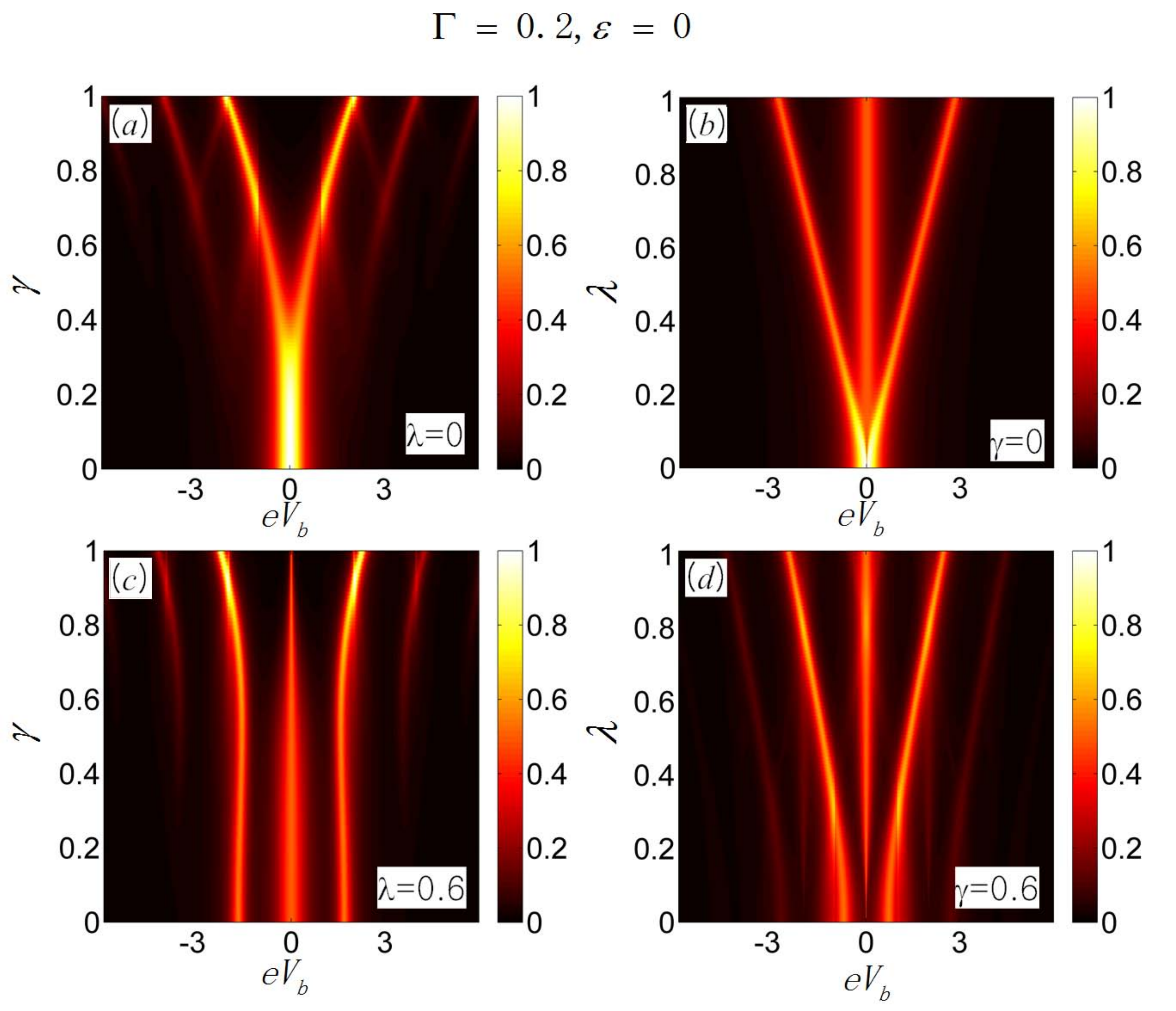}}
\caption{Spectra of the differential conductance of the single-QD system. (a) Influence of e-ph interaction on the differential conductance in the absence side-coupled MZM. (b) Conductance spectrum modified by the QD-MZM coupling without e-ph interaction. (c)-(d) Differential conductance with the interplay between the e-ph interaction and QD-MZM coupling.
\label{single}}
\end{center}
\end{figure}

\section{Numerical results and discussions \label{result2}}
\par
Based on the theory in Sec. II, we next continue to calculate the electron transport properties in our considered structure, in which the QD, suffering from the e-ph interaction, is assumed to couple laterally to one MZM. In comparison with the current through the system, the conductance spectrum is more suitable to describe the transport characteristics, so we only present the differential conductance results in the context. In this work, we would like to evaluate the influence of e-ph interaction, by considering the cases of zero and finite temperatures, respectively. The phonon frequency $\omega_0$ is taken to be the energy unit in the context\cite{can3,Zhu}.

\subsection{Results of zero-temperature limit}
In Figs.2-4, we pay attention to the differential conductance spectra in the limit of zero temperature. It is known that in this case, the phonon emission is main mechanism for the e-ph interaction\cite{Zhu}.
\par
Fig.2 shows the differential conductance results when the QD level is fixed at $\varepsilon_d=0$. In order to present the influence of e-ph interaction on the MZM-assisted transport in the QD system, we would like to first investigate the conductance spectra by ignoring the QD-MZM coupling and e-ph interaction, respectively. The MZM-absent conductance spectrum is shown in Fig.2(a), and the result without e-ph interaction is shown in Fig.2(b). It can be found in Fig.2(a) that the e-ph interaction modifies the differential conductance spectrum in two aspects. Namely, it does not only lead to the splitting of the zero-bias conductance peak but also causes new subpeaks to emerge in the nonzero-bias region. With the increase of e-ph interaction strength, the splitting of conductance peaks is enhanced, meanwhile, the conductance peaks shift to the high-bias direction. These results are not difficult to explain. The e-ph interaction enables to shift of QD level, as shown in Sec.II. Besides, it should be noticed that at the case of zero temperature, the other effect of e-ph interaction is to cause an electron to emit a phonon when it tunnels through the QD structure. On the other hand, we plot the MZM-assisted conductance spectrum in Fig.2(b), by taking the e-ph interaction out of account. It shows that the increase of QD-MZM coupling also leads to the splitting of the conductance peak, and the distances between the neighboring peaks are proportional to the QD-MZM coupling. The reason is due to that in the presence of side-coupled MZM, the conductance peaks are consistent with the eigenlevels of the QD molecule in the Nambu representation, which are $e_{1}=-\sqrt{\varepsilon_d^2+2\lambda^2}$, $e_2=0$, and $e_{3}=\sqrt{\varepsilon_d^2+2\lambda^2}$. According to the results in Eq.(12), we can easily know that the value of the zero-bias peak is equal to ${1\over2}$ (in unit of $e^2\over h$).
\par
Next, we turn to the investigation about the differential conductance properties when the MZM-assisted transport suffers from the e-ph interaction in the QD region. Fig.2(c) takes the case of $\lambda=0.6$ to show the effect of e-ph interaction. It can be found that when the e-ph interaction strength is weaker than the QD-MZM coupling, the conductance peaks are basically independent of its increment, though new subpeaks appear in the higher-bias region. Following the further increase of e-ph interaction, the zero-bias peak is narrowed to a great degree, whereas the peaks beside it are enhanced obviously, accompanied by their shift to the high-bias direction. What is notable is that the magnitude of the zero-bias conductance peak seems to be irrelevant to the existence of e-ph interaction. For comparison, we next fix the e-ph interaction and increase the QD-MZM coupling to observe the change of differential conductance, as shown in Fig.2(d). In this figure, one can find the complicated change manner of the differential conductance. Firstly, the magnitude of the zero-bias conductance peak undergoes two change processes. When $\lambda$ increases to 0.8, the zero-bias conductance peak exhibits its weak increase. Instead, the further increase of $\lambda$ only suppresses the conductance magnitude at the zero-bias limit. Thus, the zero-bias conductance peak of the MZM-assisted transport can be modified by the e-ph interaction. Such a modification depends on the interplay between the QD-MZM coupling and e-ph interaction strength. For the conductance peaks in the nonzero-bias regions, they show two maxima with the increase of $\lambda$, which appear at the points $\lambda=0.4$ and $\lambda=0.8$, respectively. Accompanied by this, the distance between these two peaks are increased. In addition, one can see that new subpeaks appear around the points of $eV_b=\pm2.0$, induced by the increase of QD-MZM coupling. They are eliminated when $\lambda>0.6$.
\begin{figure}[htb]
\begin{center}\scalebox{0.44}{\includegraphics{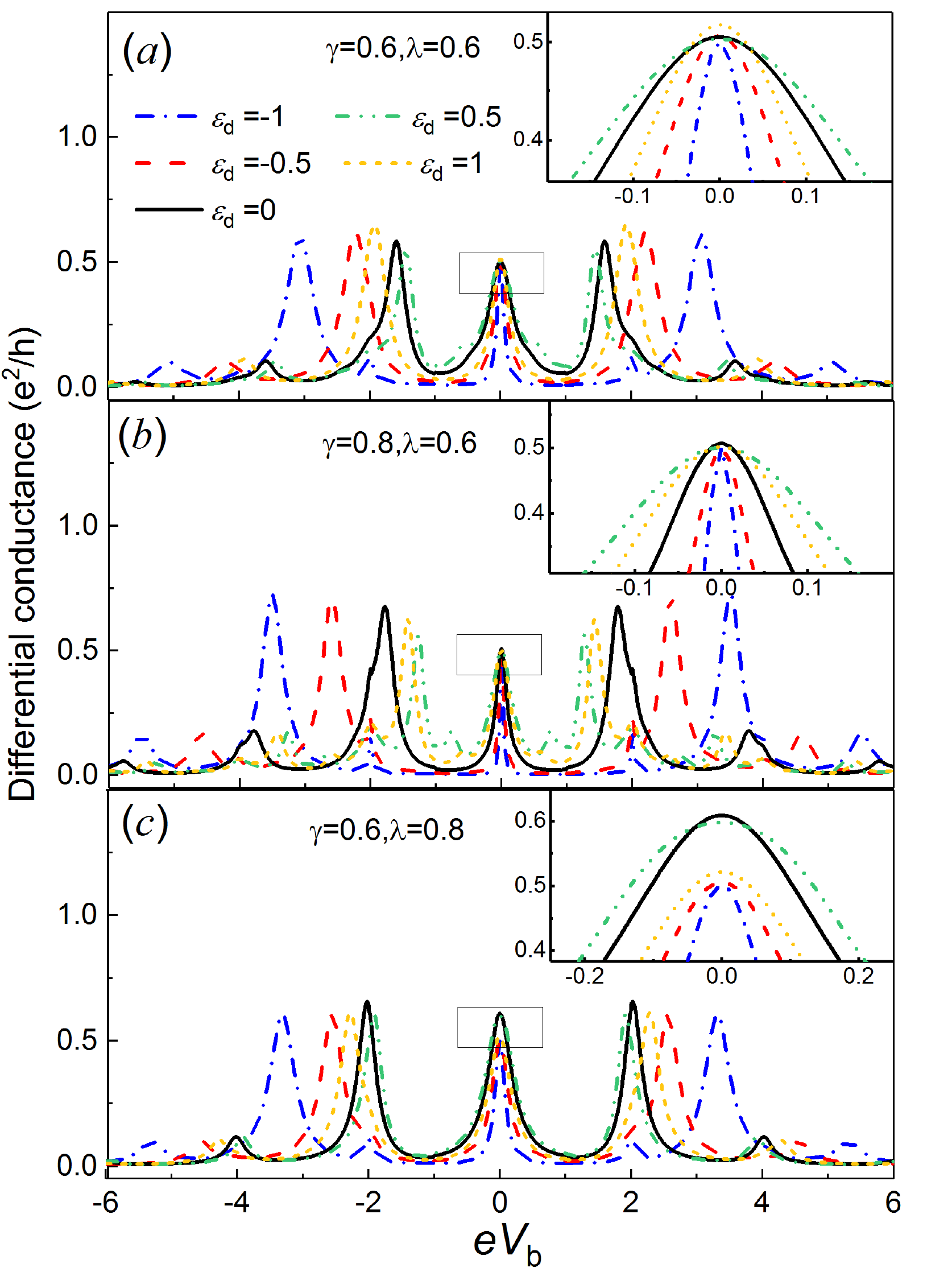}}
\caption{Effects of shifting the QD level on the differential conductance of our considered system, when the side-coupled MZM and e-p interaction coexist. In (a)-(c), the e-p interaction strength and QD-MZM coupling are considered to be $\gamma=0.6$ and $\lambda=0.6$, $\gamma=0.8$ and $\lambda=0.6$, $\gamma=0.6$ and $\lambda=0.8$, respectively.
\label{single}}
\end{center}
\end{figure}
\par
QD is characterized by the tunability of its level by applying one gate voltage. In view of this fact, we would like to adjust the QD level to study the change of the differential conductances. The numerical results are shown in Fig.3. In Fig.3(a), the e-ph interaction and QD-MZM coupling are taken to be $\gamma=\lambda=0.6$. It can be seen that when the QD level decreases from its zero value, the conductance peaks in the finite-bias region experience apparent repulsive shift. Meanwhile, the zero-bias conductance peak is narrowed very much. On the other hand, if the QD level increases from $\varepsilon_d=0.0$ to $1.0$, both the peak width and inter-peak distance change in the non-monotonous way. In the case of $\varepsilon_d=0.5$, the inter-peak distance becomes narrow and the zero-bias peak is widened. For the result of $\varepsilon_d=1.0$, the changes of the finite-bias peaks are also manifested as the enlargement of their distance and magnitude. In Fig.3(b), we plot the conductance spectra of $\gamma=0.8$ and $\lambda=0.6$. In this case, the conductance peaks are modified obviously, in comparison with Fig.3(a), due to the enhancement of e-ph interaction. For instance, when $\varepsilon_d=0.5$, new subpeaks appear in the vicinity of zero-bias point. The similarity of this two figures lies in the little change of the magnitude of the zero-bias peak. Next, the e-ph interaction is assumed to be strengthened with $\gamma=0.8$ to focus on the variation of the differential conductance, as shown in Fig.3(c). It indeed shows that in such a case, the value of the zero-bias peak almost reaches 0.6 even when $\varepsilon_d=0$. If the QD level is tuned to $\varepsilon_d=0.5$, the conductance spectrum exhibits only a little change. Nevertheless in the other cases, the magnitude of the zero-bias conductance decreases to $1\over 2$ approximately. Therefore, we know that the e-ph interaction enables to modify the zero-bias conductance peak, including its magnitude and width.
\begin{figure}[htb]
\begin{center}\scalebox{0.37}{\includegraphics{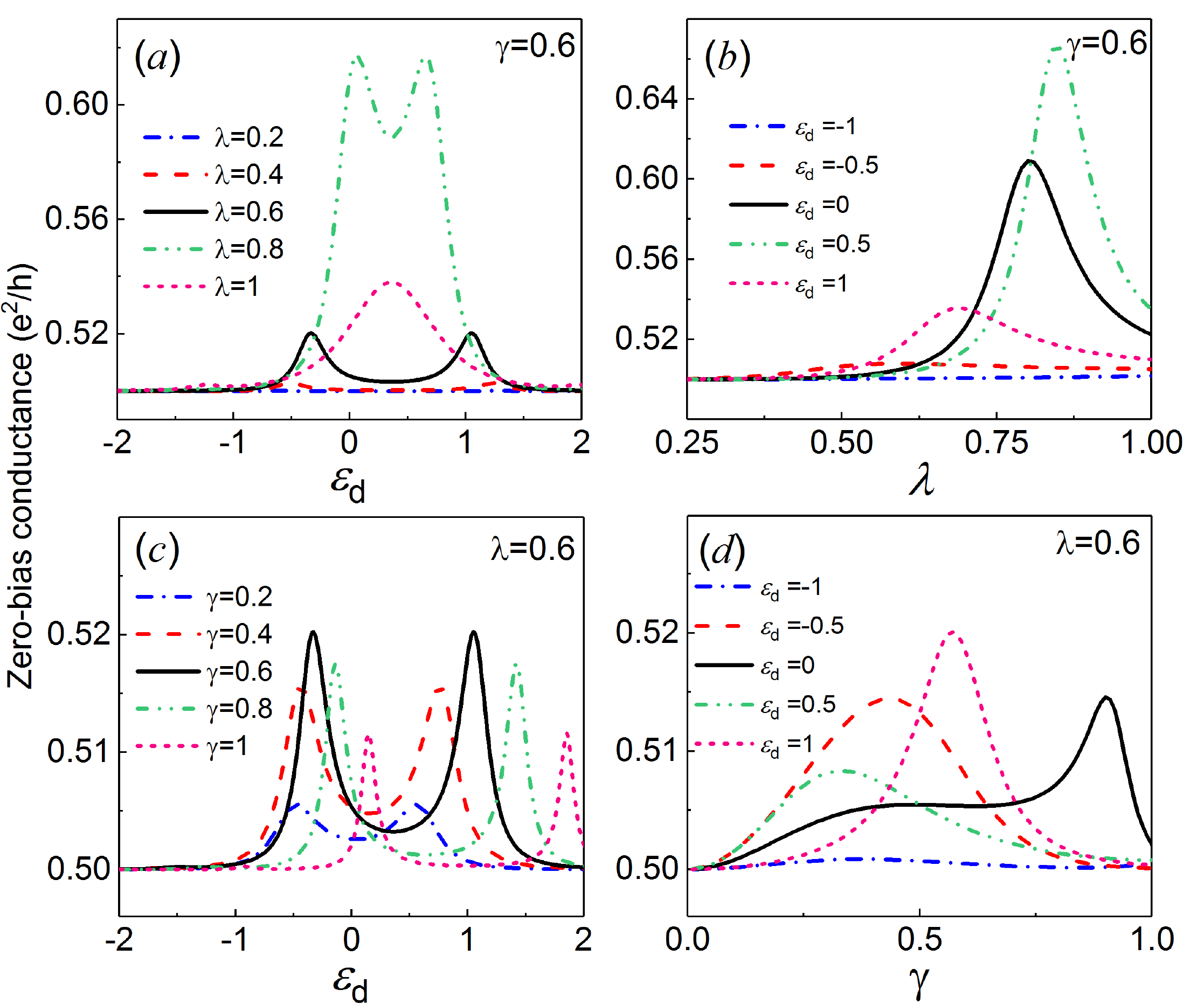}}
\caption{Results of differential conductance of our considered structure at the zero-bias limit. (a)-(b) Influences of QD level and e-ph coupling strength in the case of $\gamma=0.6$. (c)-(d) Influences of shifting the QD level by taking the QD-MZM coupling to be $\lambda=0.6$.
\label{single}}
\end{center}
\end{figure}
\par
In order to further discuss the influence of the e-ph interaction, we attempt to pay attention to the zero-bias conductance spectra in Fig.4. From this figure, one can clearly find the influence of e-ph interaction on the magnitude of the zero-bias conductance. It is certain that in the presence of e-ph interaction, the zero-bias conductance has opportunities to be greater than $1\over2$. In Fig.4(a)-(b), we take $\gamma=0.6$ and change the QD level and QD-MZM coupling, respectively. The result in Fig.4(a) shows that two additional peaks emerge in the conductance spectrum, due to presence of e-ph interaction. And the peaks are tightly related to the QD-MZM coupling. To be specific, as $\lambda$ increases to 0.6, the two additional peaks become apparent in the conductance curve, whereas the following increase of $\lambda$ can enhance these two peaks and narrows their distance. If the QD-MZM coupling is much stronger than the e-ph interaction, e.g., $\lambda=1.0$, the two peaks emerge into one, with the suppression of its magnitude. Note, also, that the e-ph interaction causes the electron-hole symmetry point to shift to the position of $\varepsilon_d={\gamma^2\over \omega_0}$, so the symmetry point of the two peaks departs from the energy zero point. Next, as shown in Fig.4(b), even in the case of $\varepsilon_d=0$, the enhancement of QD-MZM coupling is able to raise the zero-bias conductance peak until $\lambda=0.8$. The departure of the QD level to $\varepsilon_d=0.5$ can magnify the influence of e-ph interaction. Namely, it allows the zero-bias conductance reaches its maximum, about equal to 0.7.
Next, in Fig.4(c)-(d) we present the conductance spectra influenced by considering the increase of e-ph interaction. As shown in Fig.4(c), the increase of e-ph interaction enhances the conductance magnitude and enlarges the distance of the two peaks until $\gamma=0.6$. Instead when $\gamma$ further increases, the conductance peaks shift seriously to the positive-energy direction, but they are suppressed in this process. With respect to the result in Fig.4(d), it shows that when $\varepsilon_d=0$, the e-ph interaction induces the non-monotonous change. The conductance magnitude first increases slowly until $\gamma=0.7$, and then undergoes the rapid increase, with the appearance of its peak at the case of $\gamma=0.9$. When the QD level departs from its zero value, the conductance peak appears in the weaker e-p interaction case.
\par
With the help of the results above, we can clarify the effects of e-ph interaction on the conductance property of the single-QD system at the case of zero temperature. Namely, it enables to induce the increase of the zero-bias conductance magnitude. This result is completely different from the effect of e-ph interaction in the other systems. The underlying reason should be attributed to the modification of the MZM's influence on the quantum transport through this system, due to the presence of e-ph interaction.

\subsection{Finite-temperature case}
Following the results in the case of zero temperature, we continue to pay attention to the conductance properties in the finite-temperature case. In this case, the phonon emission and phonon absorption are both active, which certainly exhibit the decoherence effect.
\par
\begin{figure}[htb]
\begin{center}\scalebox{0.40}{\includegraphics{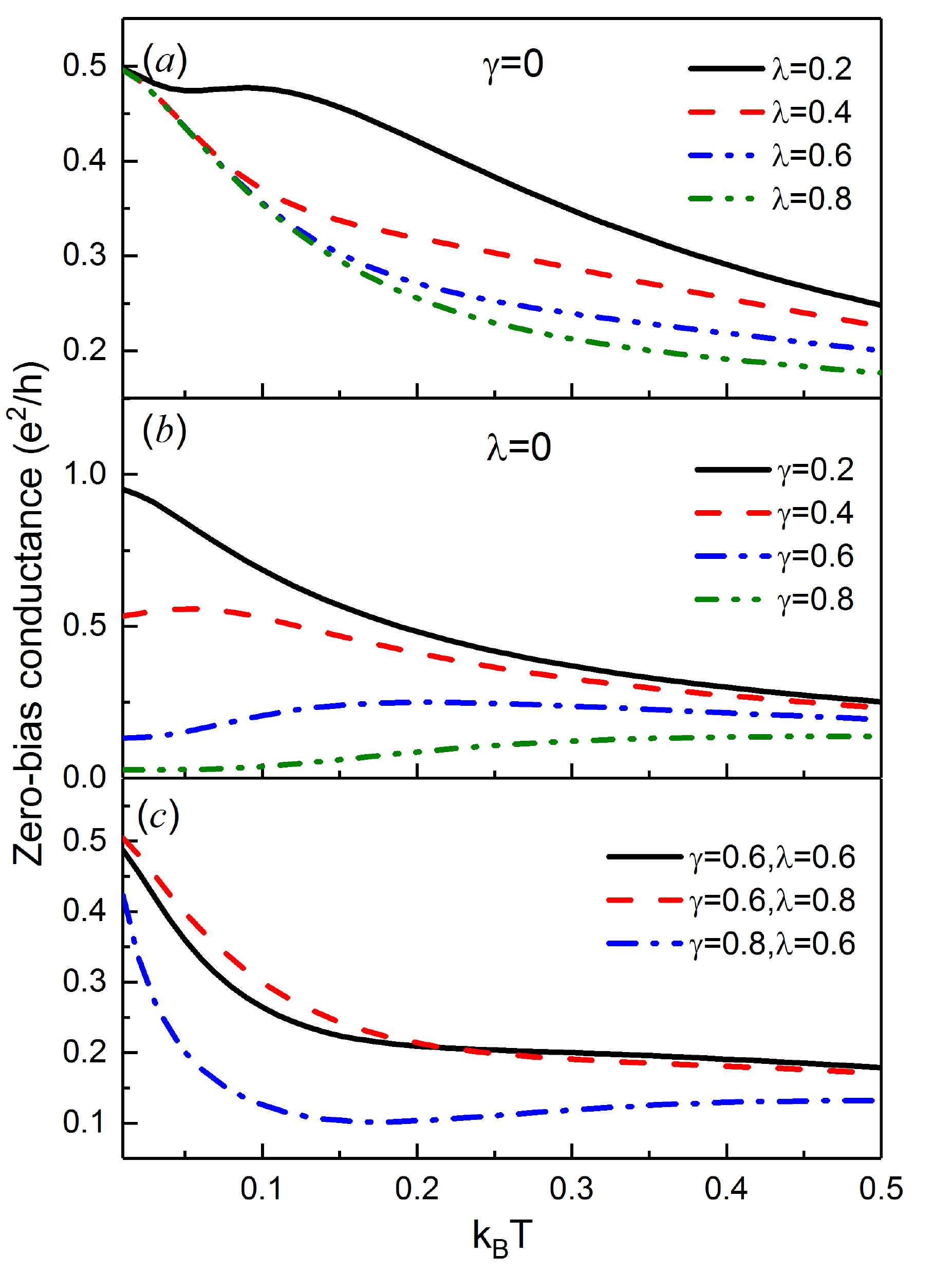}}
\caption{Zero-bias conductance with the increase of temperature. (a) Results in the absence of e-ph interaction. (b) Conductance without the QD-MZM coupling. (c) Conductance spectra due to the coexistence of e-ph and QD-MZM couplings.
\label{single}}
\end{center}
\end{figure}
Firstly, we would like to investigate the conductance variation feature at the zero-bias limit, by increasing the system's temperature. The corresponding results are shown in Fig.5. For presenting the influence of e-ph interaction on the MZM-assisted conductance properties, we plot Fig.5(a)where $\gamma=0$, Fig.5(b) where $\lambda=0$, and Fig.5(c) with $\gamma\neq0$ and $\lambda\neq0$, respectively. As shown in Fig.5(a), the conductance value is equal to 0.5 at the zero-temperature limit, in the absence of e-ph interaction. With the increase of temperature, the conductance magnitude decreases gradually, which is caused by the decoherence effect of temperature. In Fig.5(b), it can be found that for the case of single-electron tunneling, the conductance magnitude can be seriously suppressed by the increase of e-ph interaction. This indicates that the e-ph interaction mainly takes its destructive effect on the coherent transport, in the case of finite temperature. Due to this reason, the conductance suppression becomes more notable, with the increase of e-ph interaction. Next Fig.5(c) shows the results of e-ph interaction in the presence of QD-MZM coupling. One can see that in the presence of QD-MZM coupling, the e-ph interaction enables to weaken the electron transport. However, the effect is different from the case of $\lambda=0$. At the low-temperature limit, the conductance magnitude does not exhibit substantial decrease in the case of finite e-ph interaction. Based on these results, we can obtain an elementary idea about the effect of e-p interaction on the MZM-assisted transport through the QD system, in the finite-temperature case.
\par
\begin{figure}[htb]
\begin{center}\scalebox{0.37}{\includegraphics{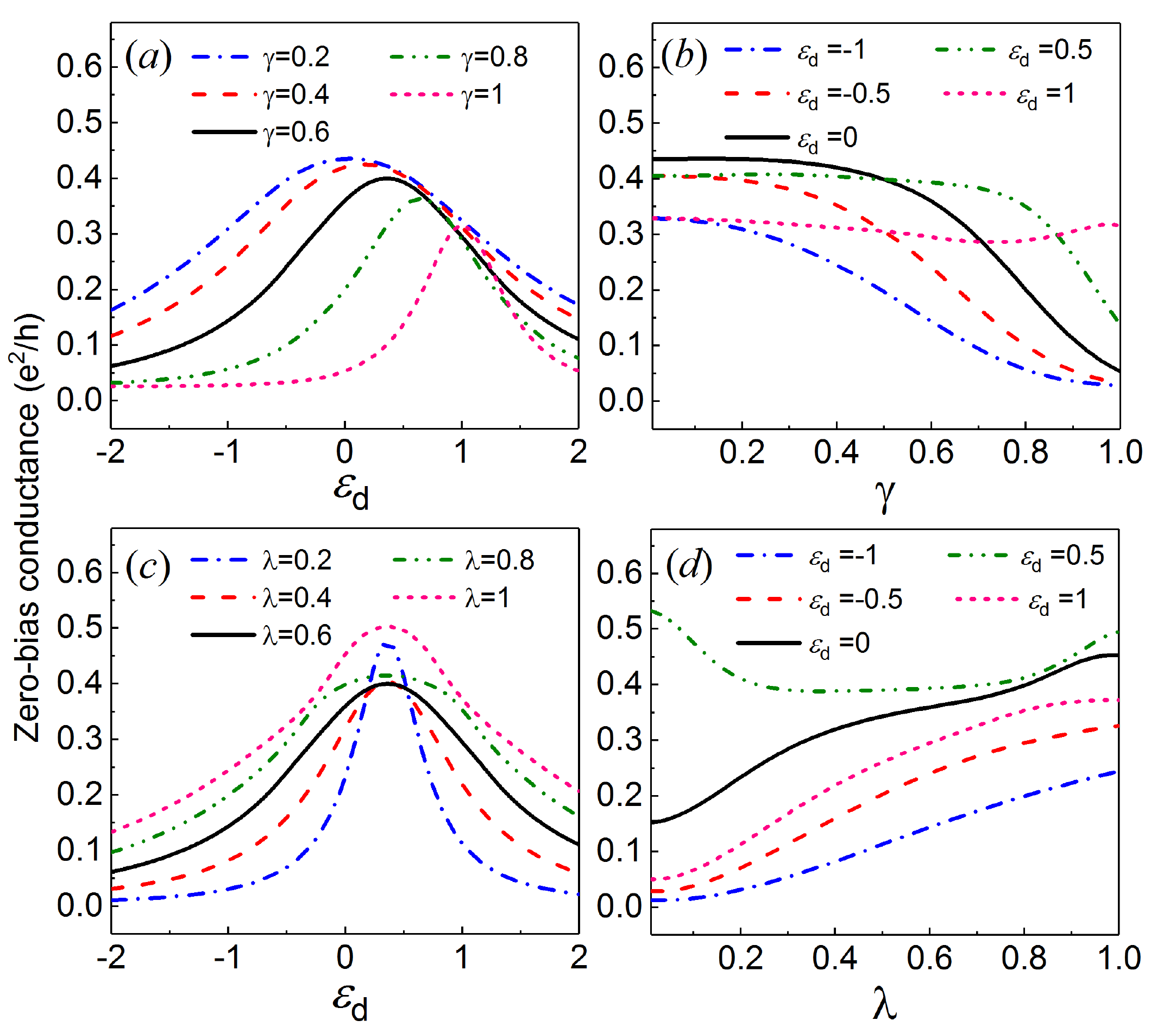}}
\caption{Zero-bias conductance in the finite-temperature case with $k_BT=0.05$. (a)-(b) Conductances curves of tuning QD level and QD-MZM coupling, by taking e-ph coupling strength to be $\lambda=0.6$. (c)-(d) Results of $\gamma=0.6$, when the QD level and QD-MZM coupling are changed.
\label{single}}
\end{center}
\end{figure}
\begin{figure}[htb]
\begin{center}\scalebox{0.45}{\includegraphics{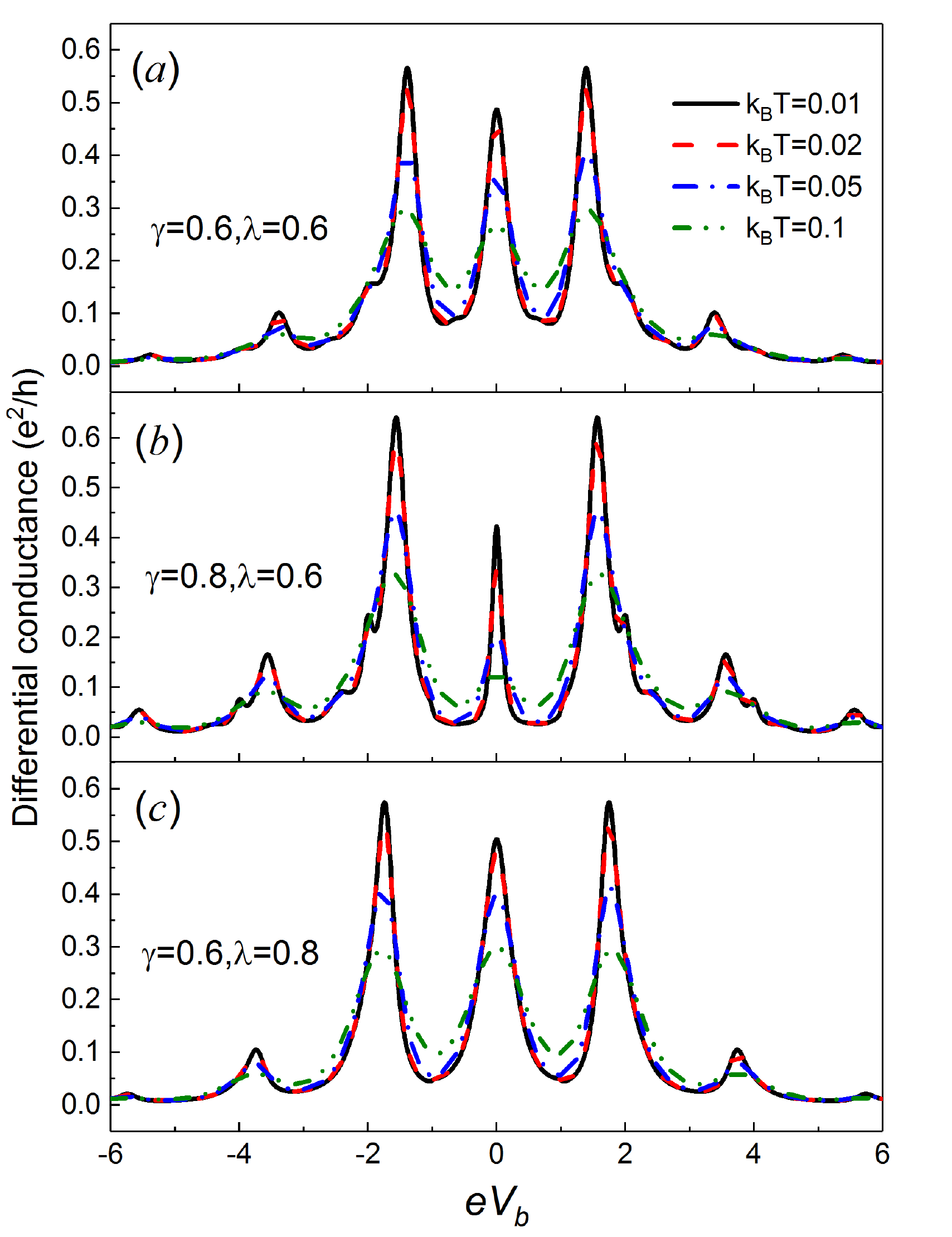}}
\caption{Differential conductance influenced by the increase of temperature. In (a)-(c), the e-ph interaction strength and QD-MZM coupling are considered to be $\gamma=0.6$ and $\lambda=0.6$, $\gamma=0.8$ and $\lambda=0.6$, $\gamma=0.6$ and $\lambda=0.8$, respectively.
\label{single}}
\end{center}
\end{figure}
\par
In Fig.6, we continue presenting the influences of e-ph interaction on the zero-bias conductance, by taking the system temperature to be $k_BT=0.05$. Fig.6(a)-(b) describe the results that the QD-MZM coupling is fixed at $\lambda=0.6$, with the adjustment of the QD level and e-ph interaction, respectively. It can be found in Fig.6(a) that even in the case of weak e-ph interaction, the conductance spectrum exhibits a Lorentian lineshape with the shift of QD level, and the conductance peak appears near the point of $\varepsilon_d=0$ with its value less than $0.5$. Once the e-ph interaction is enhanced, the conductance peak is suppressed and narrowed gradually, accompanied by its shift to the high-energy direction. This is caused by the renormalization of the QD level and QD-lead coupling due to the e-ph interaction. The results in Fig.6(b) also show the obvious effect of e-ph interaction on the zero-bias conductance. Namely, it is mainly to weaken the conductance magnitude, following the increase of e-ph interaction strength $\gamma$. The anomalous result of $\varepsilon_d=1.0$ can be explained like this. In such a case, the effective QD level has an opportunity to get close to the zero-energy point when one phonon is emitted. And then, the zero-bias transport is less dependent on the enhancement of e-ph interaction.
Next, we take $\gamma=0.6$ and plot the conductance spectra by increasing the QD-MZM coupling, as shown in Fig.6(c)-(d). One can find in Fig.6(c) that since the e-ph interaction is fixed, the QD-MZM coupling cannot move the position of the conductance peak. Instead, only its magnitude and width can be changed. In the case of $\lambda<\gamma$, the increase of QD-MZM coupling can suppress the conductance peak, accompanied by the widening of it. Nevertheless, the conductance peak can be improved by the increment of QD-MZM coupling. Next, for the role of QD-MZM coupling, we can see in Fig.6(d) that it tends to modify the decoherence effect of e-ph interaction. In the case of $\varepsilon_d=0.5$, an alternative result comes into being. This should be attributed to the non-monotonous change of the conductance peak, as shown in Fig.6(c). All these results further prove the destructive effect of e-ph interaction on the QD-assisted transport behavior.
\par

Next, Fig.7 shows the differential conductance spectra by increasing the bias voltage between the two leads. In Fig.7(a)-(c), the e-ph interaction and QD-MZM coupling are taken to be $\gamma=0.6$ and $\lambda=0.6$, $\gamma=0.8$ and $\lambda=0.6$, $\gamma=0.6$ and $\lambda=0.8$, respectively. From these three figures, we can see the suppression of the conductance magnitude with the increment of temperature, arising from the weakening of quantum coherence in this process. In addition, the interplay between the e-ph interaction and QD-MZM coupling can be clearly observed. For the stronger e-ph interaction, its-induced sub-peaks becomes more apparent. Meanwhile, the conductance peak at the low-bias limit is narrowed. As a result, the decoherence effect of temperature increase is relatively distinct, manifested as the serious suppression of the conductance peak. However, the QD-MZM coupling plays an alternative role in affecting the conductance spectrum. Interpretively, the low-bias conductance peak is wider, and the decoherence effect of temperature is weak in comparison, as shown in Fig.7(c).
\par
Up to now, one can know that at the finite-temperature case, the role of e-ph interaction is mainly to suppress the conductance magnitude of the QD system with a side-coupled MZM, since the co-existence of the phonon emission and absorption in the electron transport process.

\section{Summary}
\par
In summary, we have preformed researches about the influences of the e-ph interaction on the transport properties in one QD system with a side-coupled MZM. As a result, it has been found that at the zero-temperature limit, the zero-bias conductance value (i.e., $e^2\over 2h$) can be magnified to some extent, when the e-ph interaction is taken into account. This is manifested as the appearance of additional peaks of the zero-bias conductance. Such a phenomenon is dependent on the interplay between the e-ph interaction and QD-MZM coupling. Thus, the e-ph interaction contributes more to the MZM's influence on the quantum transport, but does not only take its decoherence effect to suppress the conductance magnitude. Alternatively, in the case of finite temperature, the e-ph interaction plays its role for the suppression of the magnitude of zero-bias conductance, but the suppression manner weaker than the case of single-electron tunneling without MZM. Therefore, one can be sure that the e-ph interaction play, we believe that this work can be helpful for further understanding the signature of the MZM in electron transport through mesoscopic circuits.
\section*{Acknowledgments}
\par
This work was financially supported by the Fundamental Research Funds for the Central Universities (Grant No. N170506007), the Liaoning BaiQianWan Talents Program (Grant No. 201892126), and the Natural Science Foundation of Shanghai (Grant No. 16ZR1447800), and the Intelligent Electronic and Systems Research Institute of Shanghai.

\clearpage
%\section{\protect\bigskip\ {\protect\large FIGURES}}

\bigskip

%\begin{figure}
%\begin{center}\scalebox{0.44}{\includegraphics{Fig1.eps}}
%\caption{ The spectra of the average electron occupation number and
%the spin accumulation in the QD. In (a)-(d) they show the influences
%of the shift of QD level on $\langle n_{\sigma}\rangle$ and $\langle
%n_{s}\rangle$. (e)-(f) $\langle n_{\sigma}\rangle$ and $\langle
%n_{s}\rangle$ vs the change of $W$ and $\phi$, respectively.
%\label{single}}
%\end{center}
%\end{figure}

%\begin{figure}
%\begin{center}\scalebox{0.40}{\includegraphics{Fig2.eps}}
%\caption{ The influences of Coulomb interaction on the properties of
%$\langle n_{\sigma}\rangle$ and $\langle n_{s}\rangle$. The relevant
%parameters are taken to be $W=0.5$, $V_\alpha=0.1$, and
%$\phi={\pi\over 2}$. \label{manybody}}
%\end{center}
%\end{figure}

\end{document}